# Analysis of Computational Science Papers from ICCS 2001-2016 using Topic Modeling and Graph Theory

Tesfamariam M. Abuhay[1], Sergey V. Kovalchuk[1], Klavdiya O. Bochenina[1],
George Kampis[1,3], Valeria V. Krzhizhanovskaya[1,2], Michael H. Lees[1,2]

[1]*ITMO University, Russia*
[2]*University of Amsterdam, The Netherlands*
[3]*Eötvös University, Hungary*
tesfaabeba@gmail.com, kovalchuk@mail.ifmo.ru, k.bochenina@gmail.com,
kampis.george@gmail.com, v.krzhizhanovskaya@uva.nl, m.h.lees@uva.nl

**Abstract**
This paper presents results of topic modeling and network models of topics using the ICCS corpus, which contains domain-specific (computational science) papers over sixteen years (a total of 5695 papers). We discuss topical structures of ICCS, how these topics evolve over time in response to the topicality of various problems, technologies and methods, and how all these topics relate to one another. This analysis illustrates multidisciplinary research and collaborations among scientific communities, by constructing static and dynamic networks from the topic modeling results and the authors' keywords. The results of this study give insights about the past and future trends of core discussion topics in computational science. We used the Non-negative Matrix Factorization(NMF) topic modeling algorithm to discover topics and labeled and grouped results hierarchically. We used Gephi to study static networks of topics, and an R library called DyA to analyze the dynamic networks of topics.

## 1 Introduction

Our aggregate knowledge continues to be digitized and filed as news, scientific articles, and books [1]. The major sources of topical scientific knowledge are conferences and scientific journals. The International Conference on Computational Science (ICCS)[*] is a yearly conference that unites specialists and researchers from computational disciplines to discuss problems and solutions in the area, to identify new issues, and to shape future directions for research. ICCS has been accepting and publishing papers of researchers from various application areas who are pioneering computational methods in sciences such as physics, chemistry, the life sciences and engineering, as well as in the arts and humanities[2–4]. Since its start in 2001, ICCS has attracted an average of 369 papers and 350 participants every year and is an A-ranked conference in CORE classification as well as one of the most cited events and publications in computational science according to Google Scholar, with an h5-index=37 and a h5-median=57.

---

[*] http://www.iccs-meeting.org/

Even though the proceedings series have turned into a noteworthy scholarly asset for computational science specialists which serves to both characterize and propel the cutting edge of the field, we believe that analyzing the underlying contents is as valuable as organizing the conference. We require automated methods to examine and comprehend this rapidly growing collection of textual data. A solution that can be used to approach this problem is *topic modeling* and the analysis of authors' keywords, as both indicate the topic area of the paper. In addition to this, we can also study the results as static and dynamic network which help us to study and understand the relationships of topics. In other words, the static and dynamic networks of topics depict multidisciplinary research and collaborations among scientific communities, where connectivity, grouping, and distances between topics and the communities of topics can be analyzed in detail.

Topic modeling is a statistical [1] and text mining based [5] method that examines the tokens of documents to detect themes or topics that run through them. The results can be used to analyze how the topics relate to one another, and how they evolve over time [1,5,6]. Topic models are based upon the idea that documents are constituted as collections of topics, where a topic is defined as a probability distribution over words [7,8]. Therefore, the main objectives of this study are to answer the following questions: *What are the underlying topical structures of ICCS papers?*, *How do these topics evolve over time?*, *Do topics relate to one another?*, *Does the relationship between topics change over time? Can we identify viable communities of topics?*

The rest of the paper is organized as follows: Section 2 describes related works and Section 3 outlines the data collection and preprocessing process. Sections 4 and 5 report static and dynamic topics of ICCS, respectively, whereas section 6 and 7 illustrate the analysis of static and dynamic networks of the topic modeling results and the authors' keywords. Finally, section 8 presents conclusion and recommendation.

## 2  Related Works

Several researchers have applied topic modeling to different kinds of corpora. For instance, Blei [7] has developed Latent Dirichlet Allocation (LDA) and illustrated the resulting model on 16,000 documents in the TREC AP corpus. Blei [1] has also applied LDA to the Yale Law Journal and 17,000 articles from Science magazine. Blei and Lafferty [9] also conducted a study on dynamic topic modeling and developed a family of probabilistic time series models to analyze the time evolution of topics and demonstrated the models by analyzing collections of the journal Science. In [8], researchers applied Latent Semantic Analysis (LSA) on the Touchstone Applied Science Associates corpus using 37,000 passages of text. Newman and Block [10] have used a probabilistic mixture decomposition method to discover topics on newspapers. Their corpus contained 80,000 articles and advertisements.

Grimmer [11] has applied a Bayesian Hierarchical Topic Model to an archive of over 24,000 press releases. More recently, a topic modeling method based on two layers of Non-negative Matrix Factorization (NMF) has been illustrated by Greene and Cross [12], who applied the method to a complete set of English language legislative speeches of the European Parliament plenary (1999-2014). In [13], the Topic over Time (TOT) model that jointly models both word co-occurrences and localization in continuous time is presented. The authors tested their model using individual emails, NIPS investigation papers and presidential state-of-the-union addresses.

However, to the best of our knowledge, no one has yet analyzed the results of topic modeling as networks of topics and studied the resulting static and dynamic connections of topics or identified communities of topics.

As mentioned above, there exist different algorithms for performing topic modeling. Of these, we have selected LDA and NMF in our experiments. NMF performs better by producing logical and distinct topics. In addition to this, researchers believe that NMF is more suitable than other topic modeling algorithms when the task is to determine both "wide" and "narrow" content with differentiated vocabularies [12,14]. Therefore, we decided to use the results of NMF. We have also utilized a log-based term frequency-inverse document frequency (TF-IDF) weighting factor to construct a document-term matrix because applying a log-based TF-IDF to topic modeling has been demonstrated as beneficial in discovering various semantically plausible topics, which are less likely to be depicted by the same high-frequency terms [12].

# 3   Data Collection and Preprocessing

We have collected all papers published in the ICCS proceedings by Springer Lecture Notes in Computer Science (LNCS) (2001-2009) and Elsevier Procedia Computer Science (2010-2016). In total, our corpus contains sixteen years of ICCS papers, the total number of papers is 5695, with each paper containing on average 10 pages of text.

The following preprocessing steps have been carried out: 1) extracting tokens using a regular expression module of Python, without punctuation marks and numbers, and changing them to lower case. 2) removing stop words such as 'the', 'and', 'or', as they contain little topical information [1]. 3) excluding words with two letters and below, and words with above thirty letters. This is because Portable Document Format (PDF) construction was done by authors in a way that made reading texts from these files difficult (the results are just sequences of long characters). Besides, according to [15], the longest valid English word contains 30 letters. 4) stemming words using Porter's stemming algorithm [16], as authors use different forms of words for grammatical reasons [1,16]. 5) constructing a document-term matrix by utilizing a log-based term frequency-inverse document frequency (TF-IDF) weighting factor.

# 4   Topical Structures of ICCS Papers

As often with clustering, we have to know the number of categories in advance. A common problem in effectively applying topic modeling methods is the choice of a proper number of topics (K). Picking too few will generate overly broad topics, while picking an excessive number will result in numerous small, highly-similar topics [12,17]. For this reason, we have conducted three preliminary experiments with the number of topics equal to 50, 100 and 150. After manually analyzing the results, we decided to fix K=100 as the topics produced captured sufficient variation while remaining semantically plausible. From this initial set of 100, we obtained 96 topics described by a sequence of word stems, ordered by their rank. Four topics were unusable as the PDF construction was done in a way that made reading texts from these files difficult. Therefore, we have excluded them from further analysis. Table 1 presents sample topic modeling results. Some words are missing endings due to the stemmer.

| | |
|---|---|
| agent, multi, action, environ, ma[†], behavior, evacu, base, social, migrat | Topic 2 |
| student, cours, teach, educ, program, project, undergradu, skill, curriculum, tool | Topic 7 |
| imag, pixel, camera, reconstruct, segment, use, process, stereo, textur, featur | Topic 10 |
| gpu, cpu, cuda, kernel, nvidia, acceler, memori, comput, devic, perform | Topic 11 |

[†] Stemmed word "MAS" (multi-agent simulation)

| | |
|---|---|
| job, schedul, execut, slot, workload, batch, shop, submiss, resourc, machin | Topic 12 |
| cell, cellular, automata, tumor, growth, tissu, biolog, immun, crypt, diffus | Topic 15 |
| packet, traffic, tcp, router, congest, transmiss, rate, queue, delay, throughput | Topic 26 |
| price, market, stock, option, trade, asset, financi, volatil, portfolio, risk | Topic 28 |
| mont, carlo, chain, random, markov, dimov, estim, stochast, invers, probabl | Topic 38 |

Table 1: Sample topic modeling results described by top ten word stems ordered by their rank

We have labeled and grouped the topic modeling results hierarchically. The hierarchy of topics was developed by a manual analysis of topics in correspondence to aims and scope of ICCS and includes three levels. The first level of hierarchy (High level, Figure 1) describes more general topics. The second level (Middle level, Figure 3:b) groups specific but similar topics to represent distinct research areas. The third level (Low level, Table 1) contains results of initial topic modeling results. In the following sections, we analyze the results in more detail. Our study indicates that the underlying topical structures or the focusing topics of ICCS are Modeling, High Performance Computing (HPC), Machine Learning, Numerical, eScience, Networks, Simulation, Data, Programming, Optimization, Visualization, Security and Education, with different proportion (Figure 1). In particular, Modeling (mainly Modeling Bioinformatics, Environmental, Multi-scale, Structural mechanics, and Agent based modeling) and High Performance Computing (mostly Parallel computing, Message Passing Interface (MPI), Fault tolerance, GPU and Cloud computing) prove to be the most interesting talking points of ICCS participants.

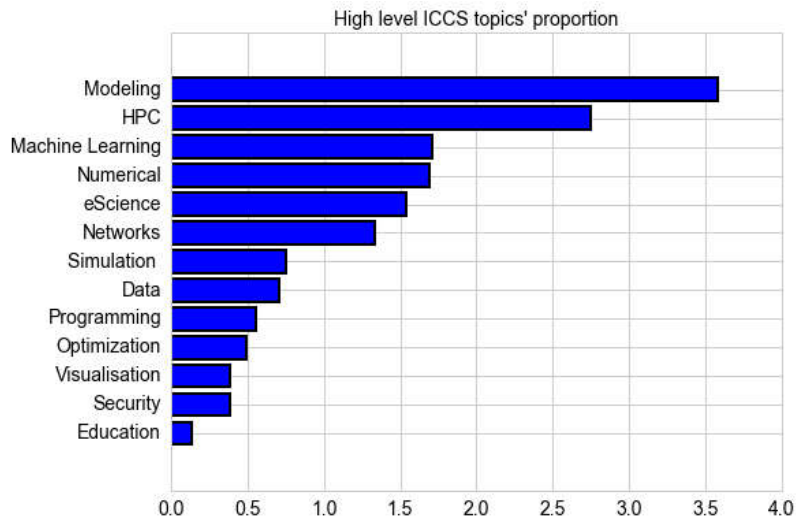

Figure 1 High-level ICCS topics' proportion in the complete dataset 2001-2016

A different way of estimating underlying topical structures of ICCS is to analyze the comparative frequency of the authors' keywords and visualize the results in a form of word cloud using an online generator such as WordClouds ‡. We have performed a preprocessing of keywords such as a replacement of abbreviations and synonym words, handling British/American writing style, and words with or without hyphens. Word clouds of the authors' keywords reveals that general topics such as Simulation, HPC, Modeling and Machine Learning are popular every year (Figure 2). The word clouds of keywords for each target year (2007-2016[§]) can be found at [19]).

---

‡ http://www.wordclouds.com/
§ In the proceedings of ICCS 2001-2006 the authors' keywords were not presented.

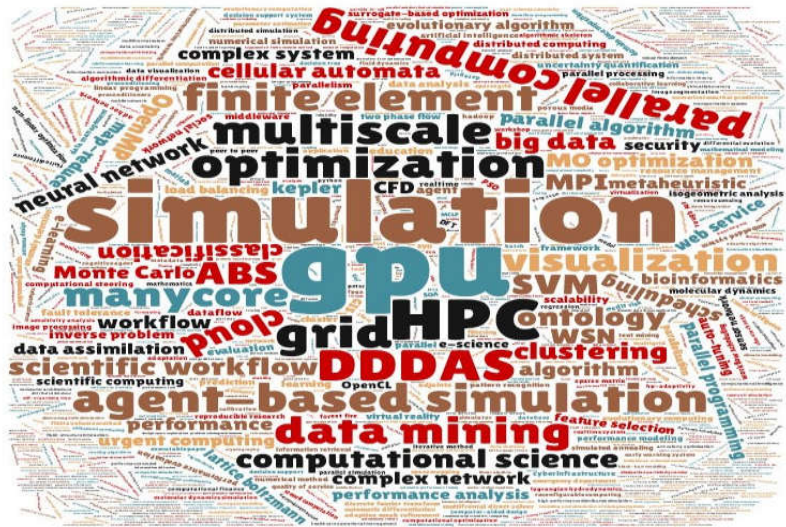

Figure 2 Word cloud for ICCS papers proceedings(2007-2016)

## 5 Evolution of ICCS Topics Over Time

Whenever a new problem area is defined, or there is a paradigm shift in the methodologies or technologies, researchers try to find a solution or change their way of doing research. This paves the way for the development of new research areas or topics, which also suggests that there is a change of research topics over time. Here we analyze how ICCS topics have evolved over the years. To do so, we collect the proportion of topics in each paper, in a year, which has contributed to the creation of a given topic. The total topic scores were then counted as a normalized sum of the collected values (which can be considered as normalized sum of topics' proportion associated with papers in one year. i.e., we have one data point in a year for one topic).

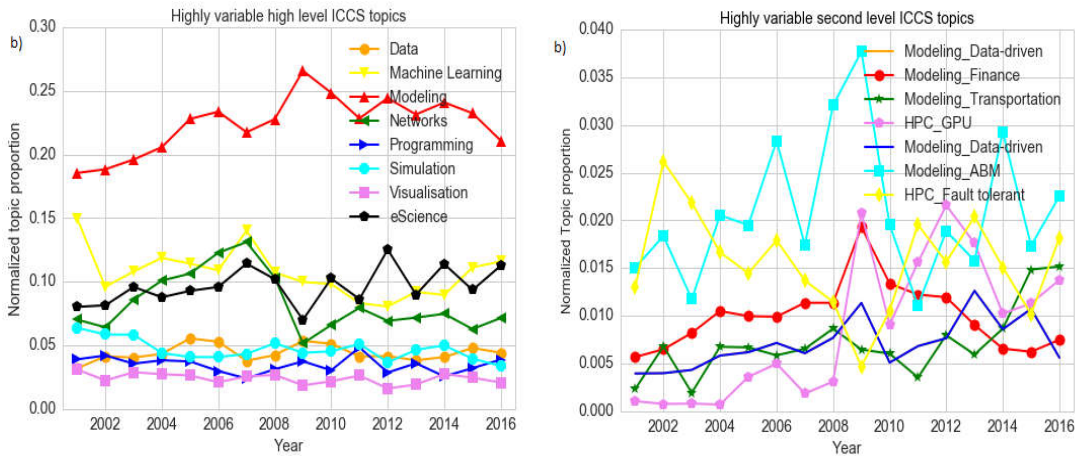

Figure 3 Highly variable high and second level ICCS topics

The analysis shows that High Performance Computing, Numerical, Security, Optimization and

Education are common high-level topics of ICCS with less variation, whereas other high-level topics show high variation over years of ICCS (Figure 3:a). For instance, "Networking" reached its peak in 2007 with significant declining after that moment, which is related to growing interest to the freshly introduced IPv6 in the computational science community.

Let us also discuss some of the highly variable second level ICCS topics and the reason behind their change over time. For instance, HPC-GPU were not popular at ICCS till 2006 (Figure 3:b). However, since 2006 (and especially, since 2009) HPC-GPU have been booming because one of the first common scientific programs which run faster on GPUs than CPUs was reported in 2005 [18], as well as appearance of GPGPU and hybrid computing libraries and standards (FireStream (2006), CUDA (2007), OpenCL (2009), and others). Another example is the topic of Modeling_Finance to which ICCS participants showed a high interest until 2009 due to the global financial crisis of 2007-2008.

We can also use Wordcloud to analyze the variability of topics over years. For example, 'executable paper' and 'reproducible research' keywords were popular in 2011. This coincides well with the fact that in 2011, there was a competition from Elsevier called "Executable paper grand challenge" [20].

## 6   Static Networks of ICCS Topics and Authors' Keywords

ICCS is one of the platforms where multidisciplinary research results have been actively reported and discussed. Next we analyze the static relationship of research topics by constructing networks of low level ICCS topics (Table 1) and a network of authors' keywords (created using all keywords from the 2007-2016 ICCS papers).

To construct the networks of topics using low level ICCS topics, first we defined three thresholds (0.025, 0.05, 0.075), marking the topics' proportion in a paper, to select those papers which have contributed more for the creation of a respective topic. Then, if two topics appear in the same paper, it means that they are connected and we used the frequency of such co-occurrences as a weight of an edge. We treated the edges as forming an undirected network. i.e., nodes representing topics, and edges representing relationships of topics with weights representing the frequency of co-occurrence of topics.

Based on prior experiments, we decided to use the threshold of 0.05 to construct networks of low level topics. As a result, we have identified six communities (where communities of topics contain related topics[**]) with an average clustering coefficient of 0.52 and an average degree of 20 - which means that one topic has a direct relationship with 20 other topics on the average. This demonstrates that these topics can be discussed together. In other words, one research topic or concept can be applied to or integrated with 20 other research topics which implies overlapping of research topics, which in return makes a demarcation of computational science research areas' difficult. We may clearly demarcate boundaries of five to six computational science research areas. This can be evidenced by an average path length which is 1.85; meaning ICCS topics are not too far from each other, theoretically or technically. The modularity is measured as 0.34.

In a network of keywords, a node represents a keyword and an edge represents a co-occurrence of two keywords in the same paper. The consolidated network of authors' keywords has 7367 nodes and 20394 edges with 448 connected components and 75.3% of nodes and 43 communities in a giant component. The average degree is equal to 5.54, showing that a keyword occurs together with 5 or 6 different keywords on average. Extremely high values of modularity (0.8) and clustering coefficient

---

[**] This may help the organizers to organize workshops based on communities of topics which will in return satisfy the interests of participants by allowing them to attend the workshop which is related to their own domain area.

(0.9) are found, explained by the principles of network creation — keywords in the same paper form a clique which has a clustering coefficient equal to 1. High clustering coefficient and small average path length (4.3) suggest that the consolidated network of authors' keywords demonstrates a small-world effect.

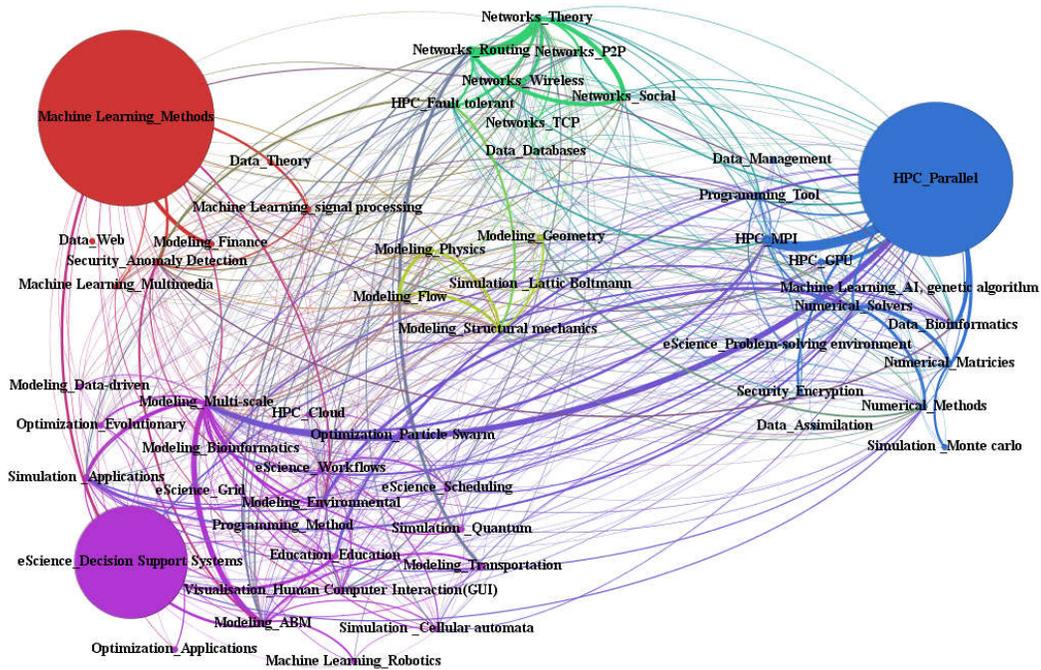

Figure 4 Networks of second level ICCS topics

To measure the relative importance of the low level ICCS topics and the authors' keywords, we calculated their centrality in the network using three different centrality metrics, namely degree, normalized betweenness and eigenvector centralities. Sorting by the value of each metric, we created three different rankings. To obtain a single ranking, we have averaged the ranks of topics, in both networks, obtained using the mentioned centrality measures (a rank is a position of a keyword in a ranking). As a result, the three metrics suggest approximately the same set of important keywords for both networks. For instance, *Modeling and Simulation* and *HPC, particularly Parallel Computing* are highly connected topics and have been serving as a hub between different research topics (see Table 2). These two topics have been playing a great role by erasing boundaries between different disciplines and facilitating multidisciplinary research.

| Networks of keywords of authors | | Networks of topic modeling results | |
|---|---|---|---|
| Average Rank | Keywords | Average Rank | Topics (word stems) |
| 1 | modeling/(and)simulation | 2.33 | Topic 85 (simul, time, model) |
| 2 | gpu/gpgpu/cuda | 2.67 | Topic 78 (parallel, processor, comput) |
| 3 | high performance computing | 3.33 | Topic 92 (memori, block, processor) |
| 5 | grid computing | 4 | Topic 2 (agent, multi, action) |
| 5.33 | optimization | 5 | Topic 3 (node, neighbor, hoc) |

Table 2 Top five low level topics and authors' keywords based on average rank

# 7 Dynamic Networks of Low Level ICCS Topics

Dynamic network analysis (DNA) can reveal additional features of topics connections over time. DNA means in our case generating a time series of static "snapshot" networks and study their evolution. We used the topics networks as defined above, and an R library called DyA [21] for their longitudinal analysis. DyA is able to produce basic time based statistics in the form of histogram series, as well as the time plots of quantities including betweenness and other centralities. The complete set, as well as a user documentation of DyA's functions can be downloaded [22,23]. Besides, DyA can analyze network evolution by percentages of replacement as networks change. The idea is also captured in the concept of "alluvial diagrams" [24]. Here we show an alluvial diagram for the last 5 years of ICCS development.[††] Note that here we study dynamic networks using the topic modeling results (i.e., Table 1).

For the betweenness centrality plot (Figure 5), we applied a threshold (of max. betweenness=400) to select low level topics with the highest values of betweenness centrality – that is, topics establishing "critical connections" with others, without which ICCS would presumably "fall apart" into separate conferences.

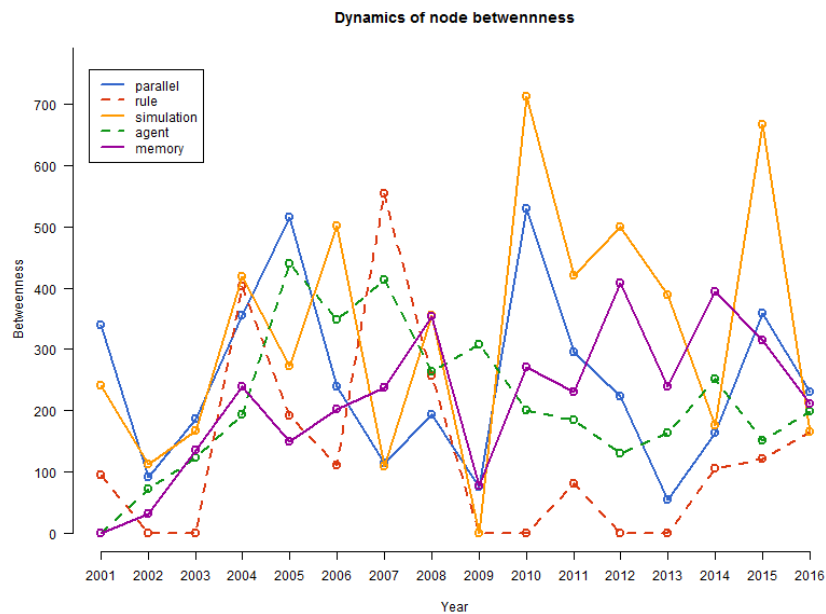

Figure 5 Highly connecting topics in the betweenness centrality plot

We can identify two noteworthy cases (Figures 5 & 6): first, the robust presence of a topic labeled as "memory", in fact marking HPC, an always present topic. This topic is also playing an important role in connecting to others. The case is different with "simulation": this topic has a less profound presence, often dissolving in other categories (Figure 6), but, the betweenness plot (Figure 5) reveals that it is nevertheless the strongest temporal connector.

---

[††] An alluvial diagram shows how network clusters change. "With clusters ordered by size, it reveals changes in network structures over time. The height of each block represents the volume of flow through the cluster, with significant subsets in [different] color." (from [24]).

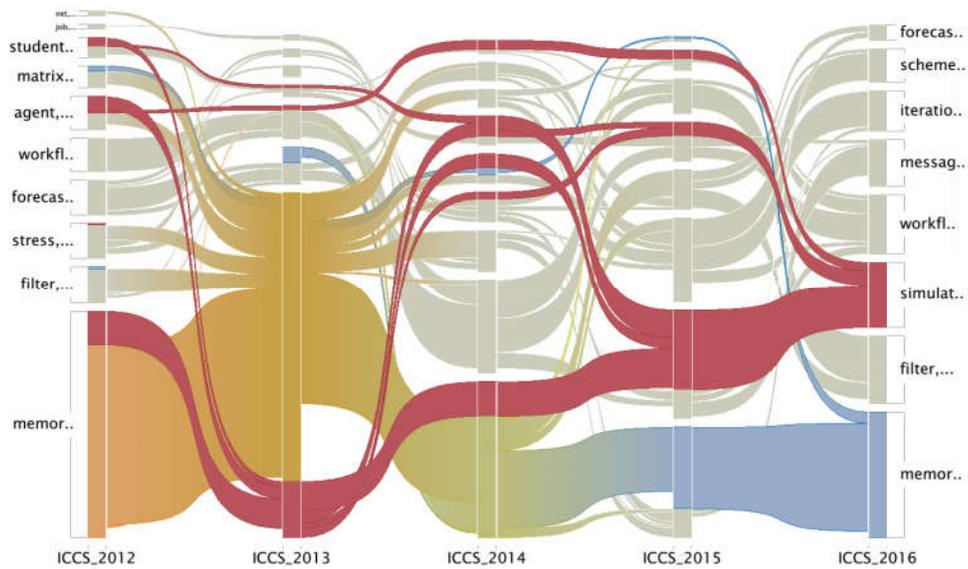

Figure 6 Transformation of the topics network for 5 years in ICCS.

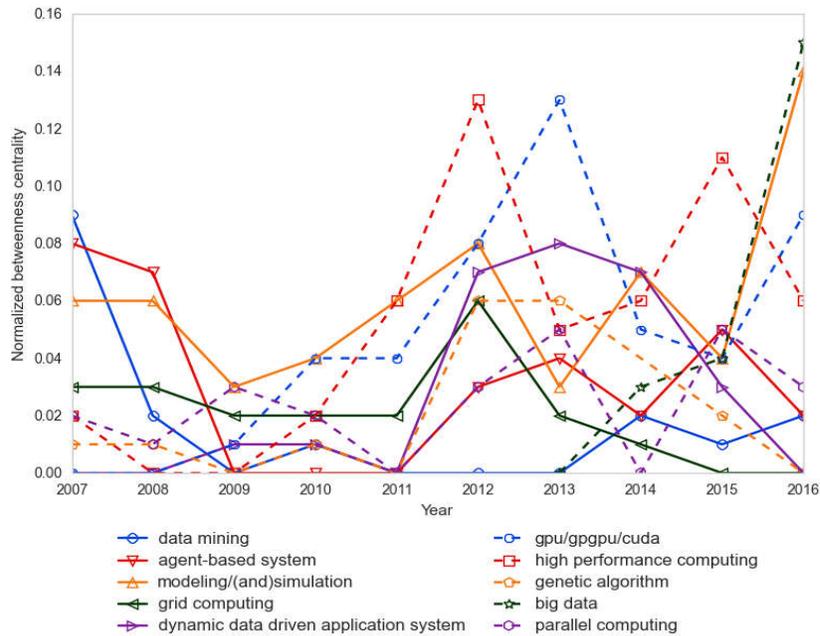

Figure 7 Dynamics of keyword betweenness centrality

In addition to this, to analyze how the importance of authors' keywords varies over time, we constructed a dynamic network of authors' keywords and plotted the dynamics of the top 10 keywords with the maximum values of betweenness centrality over the years considered. We used normalized betweenness centrality as a measure of node importance. Keywords of authors (Figure 7) can be approximately categorized into two groups: (i) permanently presented keywords with high values of betweenness centrality (e.g. again 'modeling and simulation', or 'parallel computing'), and (ii) keywords showing sharp changes indicating growth or decline of their importance. For instance, 'data mining' had the highest rank in 2007, but after that it disappears from the observation while 'big data' demonstrates a sharp rise in 2014-2016, 'DDDAS' and 'genetic algorithm' have peaks in 2012-2013, finally, 'gpu' has a constant growth from 2009 to 2013.

## 8   Conclusion and Recommendation

The main discussion topics of ICCS are Modeling, High Performance Computing (HPC), Machine Learning, Numerical, eScience, Networks, Simulation, Data, Programming, Optimization, Visualization, Security and Education, in that order with different proportion. In particular, Modeling and High Performance Computing are the most interesting discussion points of ICCS participants. However, topicality of these topics change over time due to different factors. For instance, "Networking" reached its peak in 2007 with significant declining after that moment, which is related to growing interest to introduced IPv6. Another case in point is HPC-GPU. Since 2006 (and especially, since 2009) HPC-GPU have been booming because one of the first common scientific programs which run faster on GPUs than CPUs was reported in 2005 [18], as well as appearance of GPGPU and hybrid computing libraries and standards (FireStream (2006), CUDA (2007), OpenCL (2009), and others).

At least 20 ICCS topics can be discussed together as analysis of static networks of topics demonstrates. One topic has direct relationship with 20 other topics on average. i.e., twenty topics have common characteristics which makes segregation of computational science research areas difficult. We may evidently distinguish five to six computational science research areas. This can be supported by an average path length which is 1.85; meaning ICCS topics are not too far from each other, theoretically or technically.

Keywords of author can be categorized into two. (i) static keywords representing always present topics (e.g. 'modeling and simulation', 'parallel computing'), and (ii) dynamic keywords demonstrating sharp growth or decline over years. For example, 'data mining' had the highest rank in 2007, but after that it disappears from the observation while 'big data' demonstrates sharp rise in 2014-2016, 'DDDAS' and 'genetic algorithm' reached their peaks in 2012-2013, finally, 'gpu' had constant growth from 2009 to 2013.

Simulation Modeling and High Performance Computing, particularly Parallel Computing, have been playing a great role by erasing boundaries between different disciplines and facilitating multidisciplinary research. On top of this, without these topics, ICCS would be two or more separate conferences.

Finally, even though ICCS is an A-ranked conference in CORE classification and one of the most cited events and publications in computational science according to Google Scholar, it is not the only one conference on computational science. Therefore, to make bold and concrete conclusion on the questions this study raised with respect to computational science at large, we need to also see the big picture by considering other conferences on computational science which will also give us a chance to assess the influence of other conferences on ICCS and vice versa.

*Acknowledgement.* This paper is financially supported by The Russian Scientific Foundation, Agreement #14-21-00137.